\documentclass[preprint,12pt]{elsarticle}
\usepackage{lineno}
\usepackage{graphicx}
\usepackage{bm}
\usepackage{amsmath,upgreek}
\usepackage{hyperref}
\modulolinenumbers[5]

\begin{document}

\begin{frontmatter}

\title{Statistics of bounded processes driven by Poisson white noise}
\author{S.I.\ Denisov\corref{cor1}}
\ead{denisov@sumdu.edu.ua}
\cortext[cor1]{Corresponding author}
\author{Yu.S.\ Bystrik\corref{cor2}}
\address{Sumy State University, 2 Rimsky-Korsakov Street, UA-40007
Sumy, Ukraine}

\begin{abstract}
We study the statistical properties of jump processes in a bounded domain that are driven by Poisson white noise. We derive the corresponding Kolmogorov-Feller equation and provide a general representation for its stationary solutions. Exact stationary solutions of this equation are found and analyzed in two particular cases. All our analytical findings are confirmed by numerical simulations.
\end{abstract}


\end{frontmatter}

\linenumbers

\section{Introduction}
\label{Intr}

Jump processes, non-Gaussian stochastic processes with random jumps that occur at random times, are widely used to model various phenomena in physics, chemistry, biology, economics and many other fields (see, e.g., Refs.\ \cite{VanK2007, HoLe2006, Gard2009, CoTa2004}). Often these processes are described within the continuous-time random walk (CTRW) approach \cite{MoWe_JMP1965}. Since CTRWs are characterized by only the joint distribution of jump lengths and waiting times, this approach is very simple and flexible to operate. In particular, it has been used to study the phenomena of anomalous diffusion and transport \cite{MeKl_PR2000, AvHa2005, KlRaSo2008}, superslow diffusion \cite{HaWe_JSP1990, DrKl_PRL2000, CeKlSo_EPL2003, DeKa_EPL2010} and limiting distributions of CTRWs \cite{Tun_JSP1974, ShKlWo_JSP1982, WeWeHa_JSP1989, Kot_JSP1995, MeSc_JAP2004}, including those that correspond to superheavy-tailed distributions of waiting times \cite{DeKa_PRE2011, DeYuBy_PRE2011, DeByKa_PRE2013}.

Another class of jump processes is formed by solutions of Langevin equations driven by special noises. If these noises are white, i.e., they are represented as a time derivative of stationary processes whose increments are independent on non-overlapped intervals, then solutions possess a Markovian property. It has been shown \cite{DeHoHa_EPJB2009} that the probability density functions (PDFs) of these solutions satisfy a general master equation, the so-called generalized Fokker-Planck equation. Depending on the white noise type, it reproduces the ordinary Fokker-Planck equation \cite{VanK2007, HoLe2006, Gard2009} (in the case of Gaussian white noise), the fractional Fokker-Planck equation \cite{MeKl_PR2000, KlRaSo2008} (see also Refs.\ \cite{JeMeFo_PRE1999, CGKM_ACP2006, DeHaKa_EPL2009}) (in the case of L\'{e}vy white noise), or describes the surviving and absorbing states of the system \cite{DeKaHa_JPA2010, DeKa_EPJB2011} (in the case of superheavy-tailed noise). Finally, the generalized Fokker-Planck equation that corresponds to the Langevin equation driven by Poisson white noise, i.e., a random sequence of $\delta$-pulses, is reduced to the Kolmogorov-Feller equation. In the stationary state, some exact solutions of this equation are given in Refs.\ \cite{LuBaHa_EPL1995, DaPo_PRE2006, DuRuGu_PRE2016} (see also references therein).

The Langevin equation approach is effective for studying the statistical properties of jump processes in a bounded domain as well. For example, it was used to find exact expressions for stationary PDFs in the case of L\'{e}vy white noise \cite{DeHoHa_PRE2008, DuGuBa_PRE2017}. In this paper, using the difference form of the Langevin equation with Poisson white noise, which explicitly accounts for the boundness of its solutions, we derive the corresponding Kolmogorov-Feller equation and solve it analytically in particular cases.

\section{Problem formulation}
\label{Problem}

Our aim here is to study the statistical properties of a bounded random process $X_{t}$ ($t \geq 0$) satisfying the stochastic equation
\begin{equation}
    X_{t + \tau} = S(X_{t} + \Delta_{\tau}),
    \label{eq_X}
\end{equation}
where $\tau$ is an infinitesimal time interval, $S(x)$ is the saturation function, i.e.,
\begin{equation}
    S(x) = \left\{\!\! \begin{array}{rl}
    -l, & \mathrm{if}\ x < -l,
    \\ [6pt]
    x, & \mathrm{if}\ |x| \leq l,
    \\ [6pt]
    l, & \mathrm{if}\ x > l
    \end{array}
    \right.
    \label{S(x)}
\end{equation}
($l>0$), and $\Delta_{\tau}$ is a random variable of the form
\begin{equation}
    \Delta_{\tau} = \int_{t}^{t+\tau}
    \xi(t')dt' = \int_{0}^{\tau}\xi(t')dt'.
    \label{def_Delta}
\end{equation}
Here, $\xi(t)$ is Poisson white noise (see, e.g., Ref.\ \cite{Grig2002} and references therein) defined as
\begin{equation}
    \xi(t) = \sum_{i=1}^{n(t)}z_{i}
    \delta(t-t_{i}),
    \label{def_xi}
\end{equation}
$n(t)$ is the Poisson counting process characterized by the probability $Q_{n}(t) = (\lambda t)^{n} e^{-\lambda t}/ n!$ that $n \geq 0$ events occur at random times $t_{i}$ within a given time interval $(0, t]$, $\lambda$ is the process rate parameter, $\delta(\cdot)$ is the Dirac $\delta$ function, and $z_{i}$ are independent random variables distributed with the same probability density $q(z)$, which is assumed to be symmetric. In addition, it is assumed that $\xi(t) =0$ if $n(t) =0$.

Equation (\ref{eq_X}), together with (\ref{S(x)})-(\ref{def_xi}),  shows that $X_{t}$ is a jump process whose dynamics occurs in such a way that $|X_{t}| \leq l$ for all $t \geq 0$ (we suppose that the condition $|X_{0}| \leq l$ also holds). In particular, if $X_{t}$ represents the particle coordinate, then Eq.\ (\ref{eq_X}) describes the time dependence of the position of a particle which is confined in a box (infinitely deep square well) and is driven by Poisson white noise. It should be noted that the box boundaries at $x= \pm l$ are impenetrable, i.e., $X_{t+\tau} = l$ or $-l$ if $X_{t} + \Delta_{\tau} > l$ or $X_{t} + \Delta_{\tau} < -l$, respectively.

Next, we derive the Kolmogorov-Feller equation for the PDF of the introduced process, i.e., for the probability density $P_{t}(x)$ that $X_{t} = x$.

\section{Kolmogorov-Feller equation}
\label{KFeq}

In order to find the Kolmogorov-Feller equation for $P_{t}(x)$, it is necessary first to determine the probability density $p_{\tau}(z)$ that $\Delta_{\tau} = z$, where $z \in (-\infty, \infty)$. According to (\ref{def_Delta}) and \ref{def_xi}), $\Delta_{\tau}$ is the compound Poisson process, which can be represented in the form \cite{Grig2002}
\begin{equation}
    \Delta_{\tau} =
    \left\{\!\! \begin{array}{rl}
    0, & \mathrm{if}\ n(\tau)=0,
    \\ [6pt]
    \sum_{i=1}^{n(\tau)}\!z_{i}, &
    \mathrm{if}\ n(\tau) \geq 1.
    \end{array}
    \right.
    \label{Delta}
\end{equation}
Therefore, taking into account that, by definition, $p_{\tau}(z) = \langle \delta(z - \Delta_{\tau}) \rangle$ (the angular brackets denote averaging over realizations of $\Delta_{\tau}$), the probability density $p_{\tau}(z)$ can be represented as \cite{DeHoHa_EPJB2009}
\begin{equation}
    p_{\tau}(z) = Q_{0}(\tau)\delta(z) +
    \sum_{n=1}^{\infty} Q_{n}(\tau)
    \int_{-\infty}^{\infty} \cdots
    \int_{-\infty}^{\infty}\delta
    \bigg(z - \sum_{i=1}^{n(\tau)}\!z_{i}
    \bigg) \prod_{j=1}^{n}q(z_{j})dz_{j}.
    \label{p1}
\end{equation}
Since $Q_{0}(\tau) \sim 1 - \lambda \tau$ and $Q_{n} \sim (\lambda \tau)^{n}/n!$ ($n \geq 1$) as $\tau \to 0$, from (\ref{p1}) we obtain in linear approximation in $\tau$:
\begin{equation}
    p_{\tau}(z) = (1 - \lambda \tau) \delta(z)
     + \lambda \tau q(z).
    \label{p2}
\end{equation}

As a next step, we use the definition $P_{t}(x) = \langle \delta(x - X_{t}) \rangle$ (we remind that $x \in [-l,l]$) and Eq.\ (\ref{eq_X}) to show that
\begin{eqnarray}
    P_{t+\tau}(x) \!\!\! &=& \!\!\! \langle
    \delta (x - S(X_{t} + \Delta_{\tau}))
    \rangle
    \nonumber \\[6pt]
    \!\!\! &=& \!\!\! \int_{-l}^{l} P_{t}(x')
    \bigg( \int_{-\infty}^{\infty}p_{\tau}(z)
    \delta (x - S(x' + z))dz \bigg)dx'.
    \label{P1}
\end{eqnarray}
Since
\begin{equation}
    \frac{\partial}{\partial t}P_{t}(x) =
    \lim_{\tau \to 0} \frac{1}{\tau}
    [P_{t+\tau}(x) - P_{t}(x)],
    \label{def1}
\end{equation}
from (\ref{P1}) one obtains the following integro-differential equation for $P_{t}(x)$:
\begin{equation}
    \frac{\partial}{\partial t}P_{t}(x) =
    \int_{-l}^{l}K(x,x')P_{t}(x')dx',
    \label{eq_P}
\end{equation}
where the kernel $K(x,x')$ is given by
\begin{equation}
    K(x,x') = \lim_{\tau \to 0} \frac{1}
    {\tau} \int_{-\infty}^{\infty}p_{\tau}
    (z) [\delta (x - S(x' + z)) - \delta
    (x-x')]dz.
    \label{K1}
\end{equation}

Using the definition (\ref{S(x)}) of the saturation function $S(x)$ and the normalization condition $\int_{-\infty}^{ \infty} p_{\tau} (z) dz =1$, the formula (\ref{K1}) can be rewritten as
\begin{eqnarray}
    K(x,x')\!\!\! &=& \!\!\! \delta(x-l)
    \lim_{\tau \to 0}\frac{1} {\tau}
    \int_{-x'+l}^{\infty}p_{\tau}(z)dz
    + \delta(x+l)\lim_{\tau \to 0}
    \frac{1}{\tau}\int_{-\infty}^{-x'-l}
    p_{\tau}(z)dz
    \nonumber \\[6pt]
    && \!\!\! + \lim_{\tau \to 0}\frac{1}
    {\tau}[p_{\tau}(x-x') - \delta(x-x')].
    \label{K2}
\end{eqnarray}
Then, using formula (\ref{p2}), introducing the exceedance probability
\begin{equation}
    R(x) = \int_{x}^{\infty} q(z)dz
    \label{def_R}
\end{equation}
($R(-\infty)=1$, $R(0)=1/2$, $R(\infty)=0$) and accounting for the symmetry condition $q(-z) = q(z)$, from (\ref{K2}) one straightforwardly obtains
\begin{equation}
    K(x,x') = \lambda [\delta(x-l)R(l-x')
    + \delta(x+l)R(l+x') + q(x-x') -
    \delta(x-x')].
    \label{K3}
\end{equation}
Finally, substituting this kernel into Eq.\ (\ref{eq_P}), we arrive to the following Kolmogorov-Feller equation for the bounded process $X_{t}$:
\begin{equation}
    \frac{1}{\lambda}\frac{\partial}{\partial t}
    P_{t}(x) + P_{t}(x) =
    \int_{-l}^{l}[\delta(x-l) R(l-x')
    + \delta(x+l) R(l+x') + q(x-x')]P_{t}(x')dx'.
    \label{KF}
\end{equation}
The solution of this equation must be normalized, i.e., $\int_{-l}^{l} P_{t}(x)dx = 1$, and satisfy the initial condition, which in the case of deterministic $X_{0} \in [-l,l]$ has the form $P_{0}(x) = \delta(x-X_{0})$.

Our further interest is in determining the stationary solutions $P_{\mathrm{st}}(x)$ of Eq.\ (\ref{KF}) for different $q(z)$. Since $P_{\mathrm{st}}(x) = \lim_{t \to \infty} P_{t}(x)$ and $P_{\mathrm{st}}(-x) = P_{\mathrm{st}}(x)$, this integro-differential equation reduces in the stationary state to the integral one
\begin{equation}
    P_{\mathrm{st}}(x) = [\delta(x-l) + \delta(x+l)]
    \int_{-l}^{l} R(l-x') P_{\mathrm{st}}(x')dx'
    + \int_{-l}^{l} q(x-x')P_{\mathrm{st}}(x')dx',
    \label{KFst}
\end{equation}
which does not depend on the rate parameter $\lambda$. The form of this equation suggests that its solution, the stationary PDF $P_{\mathrm{st}}(x)$, can always be represented as
\begin{equation}
    P_{\mathrm{st}}(x) = a[\delta(x-l) +
    \delta(x+l)] + f(x).
    \label{Pst}
\end{equation}
Substituting (\ref{Pst}) into Eq.\ (\ref{KFst}), it is not difficult to show that the nonnegative even function $f(x)$, which can be considered as the stationary PDF inside the bounded domain $(-l, l)$, is the solution of the Fredholm integral equation of the second type
\begin{equation}
    f(x) = a[q(x-l) + q(x+l)] + \int_{-l}^{l}
    q(x-x')f(x')dx'
    \label{f_eq}
\end{equation}
($x \in [-l,l]$) and, as it follows from the normalization condition for $P_{\mathrm{st}}(x)$, the probability $a$ that $\lim_{t \to \infty} X_{t} = l$ (or $-l$) is equal to
\begin{equation}
    a = \frac{1}{2} - \int_{0}^{l} f(x)dx.
    \label{def_a}
\end{equation}

Note also that the mean of the random process $X_{t}$ in the stationary state, $\langle X_{t} \rangle_{ \mathrm{st}} = \int_{-l}^{l} xP_{\mathrm{st}}(x) dx$, equals zero and the mean square value of $X_{t}$ in the stationary state, $\langle X_{t}^{2}\rangle_{ \mathrm{st}} = \int_{-l}^{l} x^{2}P_{ \mathrm{st}}(x) dx$, is expressed through $f(x)$ as
\begin{equation}
    \langle X_{t}^{2}\rangle_{ \mathrm{st}}
     = l^{2} - 2\int_{0}^{l}(l^{2} - x^{2})
     f(x)dx.
    \label{def_X^2}
\end{equation}

It is worthy to compare the stationary PDF (\ref{Pst}) with that for the random process $X_{t}$ driven by Gaussian white noise. In the last case, the trajectories of $X_{t}$ are continuous and $P_{t}(x)$ satisfies the ordinary Fokker-Planck equation (see, e.g., Ref.\ \cite{Gard2009})
\begin{equation}
    \frac{\partial}{\partial t}P_{t}(x)
    = D \frac{\partial^{2}}{\partial
    x^{2}}P_{t}(x),
    \label{eq_FP}
\end{equation}
where $D$ is the white noise intensity. With zero-flux boundary conditions, $\partial P_{t}(x)/\partial x|_{x= \pm l}=0$, the stationary solution of Eq.\ (\ref{eq_FP}) reads
\begin{equation}
    P_{\mathrm{st}}(x) = \frac{1}{2l}
    \quad (|x| \leq l).
    \label{Pst_FP}
\end{equation}
The same result follows also from Eq.\ (\ref{eq_P}) and the kernel representation (\ref{K2}), if the probability density $p_{\tau}(z)$ is Gaussian, i.e., if
\begin{equation}
    p_{\tau}(z) = \frac{1}{\sqrt{4\pi D\tau}}
    \exp{\!\left(- \frac{z^{2}}{4D\tau} \right)}.
    \label{p_Gauss}
\end{equation}

Comparing (\ref{Pst}) and (\ref{Pst_FP}) we see that the jumps of $X_{t}$ induced by Poisson white noise lead to nonzero probability $a$ and are responsible for nonuniformity of $f(x)$. To gain more insight into the role of these jumps, we next consider examples of $q(z)$ for which Eq.\ (\ref{f_eq}) can be solved exactly.

\section{Stationary PDFs in particular cases}
\label{St}

We assume that the probability density $q(z)$ is represented as
\begin{equation}
    q(z) = \left\{\!\! \begin{array}{ll}
    s(z), & |z| < 2l,
    \\ [6pt]
    r(z), & |z| > 2l,
    \end{array}
    \right.
    \label{def_q}
\end{equation}
where $s(-z) = s(z) \geq 0$ and $r(-z) = r(z) \geq 0$. In this case, Eq.\ (\ref{f_eq}) reduces to
\begin{equation}
    f(x) = a[s(x-l) + s(x+l)] + \int_{-l}^{l}
    s(x-x')f(x')dx'.
    \label{f_eq_b}
\end{equation}
As seen, the function $s(z)$, which is defined on the interval $(-2l, 2l)$, plays the most important role in determining $f(x)$. Since $\int_{-\infty}^{\infty} q(z)dz = 1$,  it must meet the condition
\begin{equation}
    \int_{0}^{2l} s(z)dz = \frac{1}{2}(1-\nu),
    \label{norm}
\end{equation}
where $\nu = 2 \int_{2l}^{\infty} r(z)dz$ $(0 \leq \nu \leq 1)$. We note in this context that it is not necessary to know the explicit form of the function $r(x)$ for finding $f(x)$; the only parameter $\nu$ should be specified.

Below, using special forms of the function $s(z)$ and a given value of $\nu$, we determine the stationary PDFs, i.e., calculate the function $f(x)$ and probability $a$ in two particular cases.

\subsection{First case}
\label{Case1}

Here, we assume that the function $s(z)$ satisfies the equation
\begin{equation}
    \frac{d^2}{dz^2}s(z) + \kappa^2 s(z) =0
    \label{eq_s1}
\end{equation}
($\kappa$ is a nonnegative parameter) and, as a consequence, it can be represented as
\begin{equation}
    s(z) = b \cos{(\kappa z)}.
    \label{s1}
\end{equation}
Note, since $s(z) \geq 0$ at $|z| < 2l$, the conditions $b \geq 0$ and $0 \leq 2\kappa l \leq \pi/2$ must be met.

Equation (\ref{eq_s1}) permits us to reduce the integral equation (\ref{f_eq}) to an ordinary differential equation. Indeed, by twice differentiating Eq.\ (\ref{f_eq}) with respect to $x$ and using Eq.\ (\ref{eq_s1}), one can make sure that the function $f(x)$ obeys the following ordinary differential equation:
\begin{equation}
    \frac{d^2}{dx^2}f(x) + \kappa^2 f(x) = 0.
    \label{f_eq1}
\end{equation}
Substituting its solution satisfying the symmetry condition $f(-x) = f(x)$,
\begin{equation}
    f(x) = u\cos{(\kappa x)}
    \label{sol1}
\end{equation}
($u \geq 0$, $|x| \leq l$), into Eq.\ (\ref{f_eq_b}) and using (\ref{s1}), we straightforwardly get
\begin{equation}
    u = bul + 2ab \cos{(\kappa l)} +
    \frac{bu}{2\kappa}\sin{(2\kappa l)}.
    \label{eq_abu}
\end{equation}

To determine the unknown parameters $a$, $b$ and $u$, we use Eqs.\ (\ref{def_a}), (\ref{norm}), (\ref{s1}), (\ref{sol1}) and (\ref{eq_abu}). First, from (\ref{def_a}) and (\ref{sol1}) and from (\ref{norm}) and (\ref{s1}) we find
\begin{equation}
    a = \frac{1}{2} - \frac{u}{\kappa}
    \sin{(\kappa l)}
    \label{def_a1}
\end{equation}
and
\begin{equation}
    b = \frac{(1-\nu) \kappa}{2 \sin{(2
    \kappa l)}},
    \label{b}
\end{equation}
respectively. Then, substituting (\ref{def_a1}) and (\ref{b}) into (\ref{eq_abu}), one derives
\begin{equation}
    u = \frac{2(1-\nu) \kappa \cos{
    (\kappa l)}} {(5-\nu) \sin{(2\kappa l)}
    - 2(1-\nu)\kappa l}.
    \label{u1}
\end{equation}
Finally, introducing the dimensionless function $\tilde{f} (\tilde{x}) = f(l\tilde{x})l$, variable $\tilde{x} = x/l$ ($|\tilde{x}| \leq 1$) and parameter $\rho = 2\kappa l$ ($\rho \in [0, \pi/2]$), from (\ref{sol1}), (\ref{def_a1}) and (\ref{u1}) we obtain
\begin{equation}
    \tilde{f}(\tilde{x}) = \frac{
    (1-\nu)\rho \cos{(\rho/2)}}
    {(5-\nu) \sin{\rho} - (1-\nu)
    \rho}\cos{(\rho \tilde{x}/2)}
    \label{f1}
\end{equation}
and
\begin{equation}
    a = \frac{1}{2} - \frac{(1-\nu)
    \sin{\rho}} {(5-\nu) \sin{\rho}
    - (1-\nu)\rho}.
    \label{a1}
\end{equation}

According to (\ref{f1}) and (\ref{a1}), the function $\tilde{f} (\tilde{x})$ and probability $a$ depend on the parameters $\nu$ and $\rho$. The case with $\nu = 0$ corresponds to the condition $r(z) = 0$ and the case with $\nu = 1$ to the condition $s(z) = 0$. As seen from (\ref{f1}), the function $\tilde{f} (\tilde{x})$ does not depend on $\tilde{x}$ only if $\nu = 1$ or $\rho = 0$. In the former case $\tilde{f} (\tilde{x}) =0$, and the general representation (\ref{Pst}) of the stationary PDF yields
\begin{equation}
    P_{\mathrm{st}}(x)|_{\nu=1} =
    \frac{1}{2} [\delta(x-l) +
    \delta(x+l)].
    \label{Pst1_nu=1}
\end{equation}
This formula shows that in the stationary state there are only two equiprobable values of $X_{t}$ with $X_{t}=-l$ and $X_{t}=l$. The reason is that the size of the jumps of $\Delta_{\tau}$ exceeds $2l$. Indeed, taking into account that $\int_{-\infty}^{\infty} q(z)dz = 2\int_{2l}^{\infty} r(z)dz = 1$, we have $\int_{-\infty}^ {\infty} |z|q(z)dz = 2\int_{2l}^{\infty} zr(z)dz > 2l$. In the latter case, when $\rho = 0$, from (\ref{Pst}), (\ref{f1}) and (\ref{a1}) it follows that
\begin{equation}
    P_{\mathrm{st}}(x)|_{\rho = 0} =
    \frac{1+\nu}{4} [\delta(x-l) +
    \delta(x+l)] + \frac{1-\nu}{4l}.
    \label{Pst1_rho=0}
\end{equation}

We note also that in the reference case the mean square value of $X_{t}$ in the stationary state, expression (\ref{def_X^2}), takes the form
\begin{equation}
    \langle X_{t}^{2}\rangle_{\mathrm{st}}
    = l^{2}\bigg(1 - 32\frac{(1-\nu)
    \cos{(\rho/2)}[\sin{(\rho/2)} -
    \cos{(\rho/2)}\rho/2]}
    {\rho^{2}[(5-\nu) \sin{\rho}
    - (1-\nu)\rho]} \bigg).
    \label{varX1}
\end{equation}
From this result we find $\langle X_{t}^{2}\rangle_{ \mathrm{st}} |_{\nu =1} = l^{2}$ and $\langle X_{t}^{2}\rangle_{ \mathrm{st}} |_{\rho =0} = l^{2}(2 + \nu)/3$ in accordance with (\ref{Pst1_nu=1}) and (\ref{Pst1_rho=0}), respectively.

\subsection{Second case}
\label{Case2}

If it is now assumed that
\begin{equation}
    \frac{d^2}{dz^2}s(z) - \kappa^2 s(z) =0,
    \label{eq_s2}
\end{equation}
then
\begin{equation}
    s(z) = c \cosh{(\kappa z)},
    \label{s2}
\end{equation}
where $c \geq 0$ and, contrary to the previous case, $\kappa \in [0, \infty)$. In accordance with Eq.\ (\ref{eq_s2}), in this case the integral equation (\ref{f_eq}) is reduced to
\begin{equation}
    \frac{d^2}{dx^2}f(x) - \kappa^2 f(x) = 0,
    \label{f_eq2}
\end{equation}
whose symmetric solution is given by
\begin{equation}
    f(x) = v\cosh{(\kappa x)}
    \label{sol2}
\end{equation}
($v \geq 0$). Substituting $s(z)$ and $f(x)$ from (\ref{s2}) and (\ref{sol2}) into Eq.\ (\ref{f_eq_b}), we find
\begin{equation}
    v = cvl + 2ac \cosh{(\kappa l)} +
    \frac{cv}{2\kappa}\sinh{(2\kappa l)}.
    \label{eq_acv}
\end{equation}

Proceeding in the same way as before, one obtains
\begin{equation}
    a = \frac{1}{2} - \frac{v}{\kappa}
    \sinh{(\kappa l)}
    \label{def_a2}
\end{equation}
from (\ref{def_a}) and (\ref{sol2}),
\begin{equation}
    c = \frac{(1-\nu) \kappa}{2 \sinh{(2\kappa l)}}
    \label{c}
\end{equation}
from (\ref{norm}) and (\ref{s2}), and
\begin{equation}
    v = \frac{2(1-\nu) \kappa \cosh{
    (\kappa l)}} {(5-\nu) \sinh{(2\kappa l)}
    - 2(1-\nu)\kappa l}
    \label{v}
\end{equation}
from (\ref{eq_acv}), (\ref{def_a2}) and (\ref{c}). Hence, the dimensionless function $\tilde{f}(\tilde{x})$ and the probability $a$ can be respectively written as
\begin{equation}
    \tilde{f}(\tilde{x}) = \frac{(1-\nu)
    \rho \cosh{(\rho/2)}} {(5-\nu) \sinh{
    \rho} - (1-\nu)\rho} \cosh{(\rho
    \tilde{x}/2)}
    \label{f2}
\end{equation}
($0 \leq \rho < \infty$) and
\begin{equation}
    a = \frac{1}{2} - \frac{(1-\nu)
    \sinh{\rho}} {(5-\nu) \sinh{
    \rho} - (1-\nu)\rho}.
    \label{a2}
\end{equation}

Using (\ref{f2}), (\ref{a2}) and (\ref{Pst}), it is not difficult to verify that the probability density $P_{\mathrm{st}}(x)|_{\rho = 0}$ has the same form as in (\ref{Pst1_rho=0}). At the same time, if $\rho \to \infty$ then, according to (\ref{f2}) and (\ref{a2}), we have
\begin{equation}
    \tilde{f}(\tilde{x})|_{\rho \to
    \infty} = \frac{1-\nu}{5 - \nu}
    [\delta(\tilde{x}-1) + \delta(
    \tilde{x} +1)]
    \label{f2_inf}
\end{equation}
and
\begin{equation}
    a|_{\rho \to \infty} = \frac{
    3+\nu}{2(5 - \nu)}.
    \label{a2_inf}
\end{equation}
With these results, the stationary probability density (\ref{Pst}) at $\rho \to \infty$ takes the form
\begin{equation}
    P_{\mathrm{st}}(x)|_{\rho
    \to \infty} = \frac{1}{2}
    [\delta(x-l) + \delta(x+l)],
    \label{Pst2}
\end{equation}
which, in contrast to (\ref{Pst1_nu=1}), holds for all $\nu \in [0,1]$.

Finally, using the general expression (\ref{def_X^2}) for the mean square value of the process $X_{t}$ in the stationary state and taking into account that in this case the function $f(x)$ is determined by formula (\ref{f2}), we get
\begin{equation}
    \langle X_{t}^{2}\rangle_{\mathrm{st}}
    = l^{2}\bigg(1 - 32\frac{(1-\nu)
    \cosh{(\rho/2)}[\cosh{(\rho/2)}\rho/2
    - \sinh{(\rho/2)}]}
    {\rho^{2}[(5-\nu) \sinh{\rho}
    - (1-\nu)\rho]} \bigg).
    \label{varX2}
\end{equation}
From this it follows that, as in the previous case, $\langle X_{t}^{2}\rangle_{ \mathrm{st}}|_{\rho \to 0} = l^{2}(2 + \nu)/3$ and, in accordance with (\ref{Pst2}), $\langle X_{t}^{2}\rangle_{ \mathrm{st}}|_{\rho \to \infty} = l^{2}$.

\section{Numerical results}
\label{Num}

To verify our analytical results, we solved Eq.\ (\ref{eq_X}) numerically $N$ times assuming that $\tau$ is the time step size and $X_{0} =0$. For calculating $f(x)$ and $a$ in the above considered cases we proceed as follows. First, we divide the interval $(-l, l)$ into $K$ subintervals of length $\delta = 2l/K$ and introduce the left and right coordinates of the $k$th subinterval as $x_{k}^{-} = -l + (k-1)\delta$ and $x_{k}^{+} = -l + k\delta$, respectively, where $k = \overline{1, K}$. Then the probability density that $X_{M\tau} = x_{k} = (x_{k}^{-} + x_{k}^{+})/2$ (the number of time steps $M$ must be large enough to reach the stationary state) can be approximately defined as $f(x_{k}) = N_{k}/ 2lN$. Here, $N_{k}$ is the number of runs for which $X_{M\tau} \in [x_{k}^{-}, x_{k}^{+})$ and $k = \overline{2,K-1}$. Similarly, the probabilities that $X_{M\tau}$ belongs to the leftmost and rightmost subintervals $[-l, -l+\delta)$ and $[l - \delta, l]$ are defined as $a_{1} = N_{1}/N$ and $a_{K} = N_{K}/N$, respectively.

\begin{figure}[ht]
    \centering
    \includegraphics[width=\columnwidth]{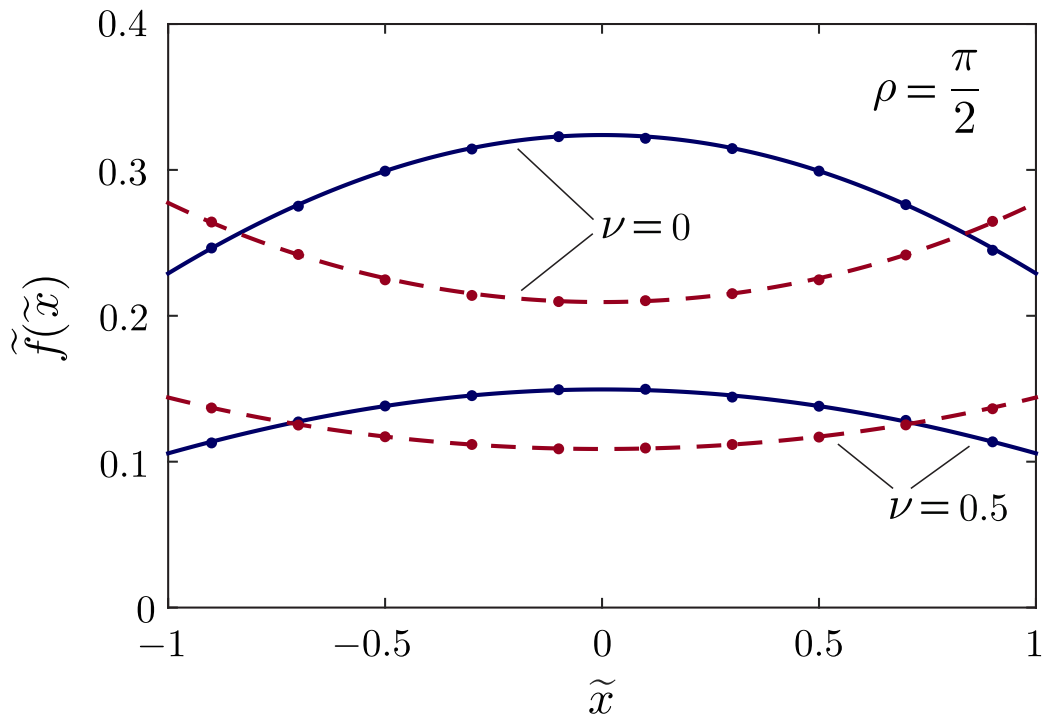}
    \caption{Plots of the function $\tilde{f}
    (\tilde{x})$ for $\rho = \pi/2$ and different
    values of the probability $\nu$. The solid and
    dashed curves correspond to analytical results
    (\ref{f1}) and (\ref{f2}), respectively. The
    numerical results obtained by simulating Eq.\
    (\ref{eq_X}) are denoted by circle symbols.}
    \label{fig1}
\end{figure}

\begin{figure}[ht]
    \centering
    \includegraphics[width=\columnwidth]{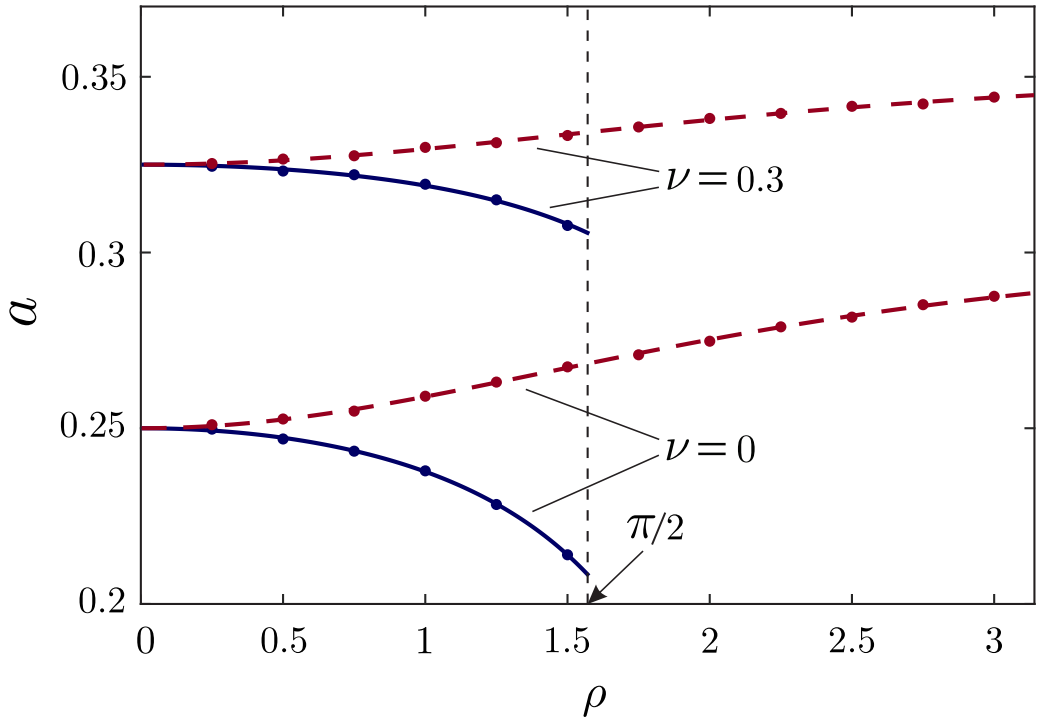}
    \caption{The probability $a$ as a function of
    the parameter $\rho$. The solid and dashed
    curves correspond to analytical results
    (\ref{a1}) and (\ref{a2}), respectively, and the
    numerical results are denoted by circle symbols.}
    \label{fig2}
\end{figure}

The analytical results for $\tilde{f}(\tilde{x})$ and the results of numerical simulations of $\tilde{f} (\tilde{x}_{k})$ are shown in Fig.\ \ref{fig1} for both considered cases (the other simulation parameters are chosen to be $\lambda \tau = 10^{-2}$, $N= 10^{6}$, $K= 10$ and $M= 10^{4}$). As seen, the analytical results (\ref{f1}) and (\ref{f2}) are in excellent agreement with numerical ones. Note also that according to our predictions the values of $\tilde{f} (\tilde{x}_{k})$ depend only on the probability $\nu$ and not on the explicit form of the function $r(z)$ in the representation (\ref{def_q}). As shown in Fig.\ \ref{fig2}, the theoretical results (\ref{a1}) and (\ref{a2}) for the probability $a$ are also in fair agreement with numerical simulations ($a = a_{1} = a_{K}$).

\section{Conclusions}
\label{Concl}

We have studied the statistical properties of a class of random processes that are bounded by the saturation function and are driven by Poisson white noise. We have derived the Kolmogorov-Feller equation for the probability density function (PDF) of these processes and found a general representation of its solution in the stationary state. This representation has two important features arising from both processes jumps and their boundedness. One of them is that the stationary PDF inside the bounded domain is, in general, nonuniform and the other is that the probability of extremal values of these processes is nonzero. It has been shown that this stationary PDF satisfies the Fredholm integral equation of the second type and uniquely determines the mentioned probability. We have solved the Fredholm equation analytically in two particular cases, calculated the probabilities of extremal values of the reference processes and determined their mean square values.

In view of these observations it seems that the bounded processes driven by Poisson white noise can be used to model the destruction  processes in different systems. In this approach, the probability of extremal values of the state variable (e.g., in biological and technical systems) can be associated with the destruction probability.

\section*{Acknowledgments}

This work was supported by the Ministry of Education and Science of Ukraine under Grant No.\ 0116U002622.


\section*{References}

\begin{thebibliography} {99}

\bibitem{VanK2007} N.G.\ van Kampen, Stochastic Processes in Physics and Chemistry, 3rd ed., Elsevier, North Holland, 2007. ISBN: 978-0-444-52965-7.
\bibitem{HoLe2006} W.\ Horsthemke, R.\ Lefever, Noise-Induced Transitions. Theory and Applications in Physics, Chemistry, and Biology, 2nd pr., Springer-Verlag, Berlin, 2006. ISBN: 978-3-540-11359-1.
\bibitem{Gard2009} C.W.\ Gardiner, Stochastic Methods: A Handbook for the Natural and Social Sciences, 4th ed., Springer-Verlag, Berlin, 2009. ISBN: 978-3-540-70712-7.
\bibitem{CoTa2004} R.\ Cont, P.\ Tankov, Financial Modeling with Jump Processes, Chapman \& Hall/CRC, Boca Raton, 2004. ISBN:  1-58488-413-4.

\bibitem{MoWe_JMP1965} E.W.\ Montroll, G.H.\ Weiss, Random walks on lattices. II, J.\ Math.\ Phys.\ 6 (1965) 167-181.
    \newblock \href {http://dx.doi.org/10.1063/1.1704269}
    {\path{doi:10.1063/1.1704269}}.

\bibitem{MeKl_PR2000} R.\ Metzler, J.\ Klafter, The random walk's guide to anomalous diffusion: a fractional dynamics approach Phys.\ Rep.\ 339 (2000) 1-77.
    \newblock \href {http://dx.doi.org/10.1016/S0370-1573(00)00070-3}
    {\path{doi:10.1016/S0370-1573(00)00070-3}}.
\bibitem{AvHa2005} D.\ ben-Avraham, S.\ Havlin, Diffusion and Reactions in Fractals and Disordered Systems, Cambridge University Press, Cambridge, 2005. ISBN: 0-521-61720-0.
\bibitem{KlRaSo2008} R.\ Klages, G.\ Radons, I.M.\ Sokolov (Eds.), Anomalous Transport: Foundations and Applications, Wiley-VCH, Berlin, 2008. ISBN: 978-3-527-40722-4.

\bibitem{HaWe_JSP1990} S.\ Havlin, G.H.\ Weiss, A new class of long-tailed pausing time densities for the CTRW, J.\ Stat.\ Phys.\ 58 (1990) 1267-1273.
    \newblock \href {http://dx.doi.org/10.1007/BF01026577}
    {\path{doi:10.1007/BF01026577}}.
\bibitem{DrKl_PRL2000} J.\ Dr\"{a}ger, J.\ Klafter, Strong anomaly in diffusion generated by iterated maps, Phys.\ Rev.\ Lett.\ 84 (2000) 5998-6001.
    \newblock \href {http://dx.doi.org/10.1103/PhysRevLett.84.5998}
    {\path{doi:10.1103/PhysRevLett.84.5998}}.
\bibitem{CeKlSo_EPL2003} A.V.\ Chechkin, J.\ Klafter, I.M.\ Sokolov, Fractional Fokker-Planck equation for ultraslow kinetics,  Europhys.\ Lett.\ 63 (2003) 326-332.
    \newblock \href {http://dx.doi.org/10.1209/epl/i2003-00539-0}
    {\path{doi:10.1209/epl/i2003-00539-0}}.
\bibitem{DeKa_EPL2010} S.I.\ Denisov, H.\ Kantz, Continuous-time random walk theory of superslow diffusion, Europhys.\ Lett.\ 92 (2010) 30001.
    \newblock \href {http://dx.doi.org/10.1209/0295-5075/92/30001}
    {\path{doi:10.1209/0295-5075/92/30001}}.

\bibitem{Tun_JSP1974} J.K.E.\ Tunaley, Asymptotic solutions of the continuous-time random walk model of diffusion, J.\ Stat.\ Phys.\ 11 (1974) 397-408.
    \newblock \href {http://dx.doi.org/10.1007/BF01026731}
    {\path{doi:10.1007/BF01026731}}.
\bibitem{ShKlWo_JSP1982} M.F.\ Shlesinger, J.\ Klafter, Y.M.\ Wong, Random walks with infinite spatial and temporal moments, J.\ Stat.\ Phys.\ 27 (1982) 499-512.
    \newblock \href {http://dx.doi.org/10.1007/BF01011089}
    {\path{doi:10.1007/BF01011089}}.
\bibitem{WeWeHa_JSP1989} H.\ Weissman, G.H.\ Weiss, S.\ Havlin, Transport properties of the continuous-time random walk with a long-tailed waiting-time density, J.\ Stat.\ Phys.\ 57 (1989) 301-317.
    \newblock \href {http://dx.doi.org/10.1007/BF01023645}
    {\path{doi:10.1007/BF01023645}}.
\bibitem{Kot_JSP1995} M.\ Kotulski, Asymptotic distributions of continuous-time random walks: A probabilistic approach, J.\ Stat.\ Phys.\ 81 (1995) 777-792.
    \newblock \href {http://dx.doi.org/10.1007/BF02179257}
    {\path{doi:10.1007/BF02179257}}.
\bibitem{MeSc_JAP2004} M.M.\ Meerschaert, H.-P.\ Scheffler, Limit theorems for continuous-time random walks with infinite mean waiting times, J.\ Appl.\ Prob.\ 41 (2004) 623-638.
    \newblock \href {http://dx.doi.org/10.1239/jap/1091543414}
    {\path{doi:10.1239/jap/1091543414}}.

\bibitem{DeKa_PRE2011} S.I.\ Denisov, H.\ Kantz, Continuous-time random walk with a su\-per\-heavy-tailed distribution of waiting times, Phys.\ Rev.\ E 83 (2011) 041132.
    \newblock \href {http://dx.doi.org/10.1103/PhysRevE.83.041132}
    {\path{doi:10.1103/PhysRevE.83.041132}}.
\bibitem{DeYuBy_PRE2011} S.I.\ Denisov, S.B.\ Yuste, Yu.S.\ Bystrik, H.\ Kantz, K.\ Lindenberg, Asymptotic solutions of decoupled continuous-time random walks with superheavy-tailed waiting time and heavy-tailed jump length distributions, Phys.\ Rev.\ E 84 (2011) 061143.
    \newblock \href {http://dx.doi.org/10.1103/PhysRevE.84.061143}
    {\path{doi:10.1103/PhysRevE.84.061143}}.
\bibitem{DeByKa_PRE2013} S.I.\ Denisov, Yu.S.\ Bystrik, H.\ Kantz, Limiting distributions of con\-tinuous-time random walks with superheavy-tailed waiting times, Phys.\ Rev.\ E 87 (2013) 022117.
    \newblock \href {http://dx.doi.org/10.1103/PhysRevE.87.022117}
    {\path{doi:10.1103/PhysRevE.87.022117}}.

\bibitem{DeHoHa_EPJB2009} S.I.\ Denisov, W.\ Horsthemke, P.\ H\"{a}nggi, Generalized Fokker-Planck equation: Derivation and exact solutions, Eur.\ Phys.\ J.\ B 68 (2009) 567–575.
    \newblock \href {http://dx.doi.org/10.1140/epjb/e2009-00126-3}
    {\path{doi:10.1140/epjb/e2009-00126-3}}.

\bibitem{JeMeFo_PRE1999} S.\ Jespersen, R.\ Metzler, H.C.\ Fogedby, L\'{e}vy flights in external force fields: Langevin and fractional Fokker-Planck equations and their solutions, Phys.\ Rev.\ E 59 (1999) 2736-2745.
    \newblock \href {http://dx.doi.org/10.1103/PhysRevE.59.2736}
    {\path{doi:10.1103/PhysRevE.59.2736}}.
\bibitem{CGKM_ACP2006} A.V.\ Chechkin, V.Y.\ Gonchar, J.\ Klafter, R.\ Metzler, Fundamentals of L\'{e}vy flight processes, Adv.\ Chem.\ Phys.\ 133 (2006) 439-496.
    \newblock \href {http://dx.doi.org/10.1002/0470037148.ch9}
    {\path{doi:10.1002/0470037148.ch9}}.
\bibitem{DeHaKa_EPL2009} S.I.\ Denisov, P.\ H\"{a}nggi, H.\ Kantz, Parameters of the fractional Fokker-Planck equation, Europhys.\ Lett.\ 85 (2009) 40007.
    \newblock \href {http://dx.doi.org/10.1209/0295-5075/85/40007}
    {\path{doi:10.1209/0295-5075/85/40007}}.

\bibitem{DeKaHa_JPA2010} S.I.\ Denisov, H.\ Kantz, P.\ H\"{a}nggi, Langevin equation with super-heavy-tailed noise, J.\ Phys.\ A: Math.\ Theor.\ 43 (2010) 285004.
    \newblock \href {http://dx.doi.org/10.1088/1751-8113/43/28/285004}
    {\path{doi:10.1088/1751-8113/43/28/285004}}.
\bibitem{DeKa_EPJB2011} S.I.\ Denisov, H.\ Kantz, Probability distribution function for systems driven by superheavy-tailed noise, Eur.\ Phys.\ J.\ B 80 (2011) 167–175.
    \newblock \href {http://dx.doi.org/10.1140/epjb/e2011-10758-1}
    {\path{doi:10.1140/epjb/e2011-10758-1}}.

\bibitem{LuBaHa_EPL1995} J.\ {\L}uczka, R.\ Bartussek, P.\ H\"{a}nggi, White-noise-induced transport in periodic structures, Europhys.\ Lett.\ 31 (1995) 431-436.
    \newblock \href {http://dx.doi.org/10.1209/0295-5075/31/8/002}
    {\path{doi:10.1209/0295-5075/31/8/002}}.
\bibitem{DaPo_PRE2006} E.\ Daly, A.\ Porporato, Probabilistic dynamics of some jump-diffusion systems, Phys.\ Rev.\ E 73 (2006) 026108.
    \newblock \href {http://dx.doi.org/10.1103/PhysRevE.73.026108}
    {\path{doi:10.1103/PhysRevE.73.026108}}.
\bibitem{DuRuGu_PRE2016} A.A.\ Dubkov, O.V.\ Rudenko, S.N.\ Gurbatov, Probability characteristics of nonlinear dynamical systems driven by $\delta$-pulse noise, Phys.\ Rev.\ E 93 (2016) 062125.
    \newblock \href {http://dx.doi.org/10.1103/PhysRevE.93.062125}
    {\path{doi:10.1103/PhysRevE.93.062125}}.

\bibitem{DeHoHa_PRE2008} S.I.\ Denisov, W.\ Horsthemke, P.\ H\"{a}nggi, Steady-state L\'{e}vy flights in a confined domain, Phys.\ Rev.\ E 77 (2008) 061112.
    \newblock \href {http://dx.doi.org/10.1103/PhysRevE.77.061112}
    {\path{doi:10.1103/PhysRevE.77.061112}}.
\bibitem{DuGuBa_PRE2017} B.\ Dybiec, E.\ Gudowska-Nowak, E.\ Barkai, A.A.\ Dubkov, L\'{e}vy flights versus L\'{e}vy walks in bounded domains, Phys.\ Rev.\ E 95 (2017) 052102.
    \newblock \href {http://dx.doi.org/10.1103/PhysRevE.95.052102}
    {\path{doi:10.1103/PhysRevE.95.052102}}.

\bibitem{Grig2002} M.\ Grigoriu,  Stochastic Calculus: Applications in Science and Engineering, Birkh\"{a}user, Boston, 2002.    ISBN: 978-0-8176-8228-6.

\end{thebibliography}
\end{document}